\documentclass{elsart}
\usepackage[dvips]{graphicx}
\usepackage{dcolumn}
\usepackage{amsmath}
\usepackage{amssymb}
\usepackage{bm}
\journal{Chemical Physics Letters}

\begin{document}
\begin{frontmatter}

\title{A Proposal for a Modified M{\o}ller-Plesset Perturbation Theory}
\author[acabo]{Alejandro Cabo},
\author[fclaro]{Francisco Claro}, and
\author[emenendez]{Eduardo Men\'endez-Proupin\corauthref{cor}}
\corauth[cor]{Corresponding author. E-mail: emenendez@macul.ciencias.uchile.cl}

\address[acabo]{Grupo de F\'{i}sica Te\'orica, Instituto de Cibern\'etica,
Matematica y F\'{i}sica, Calle E, No. 309, Vedado, La Habana, Cuba }
\address[fclaro]{Facultad de F\'{\i}sica, Pontificia Universidad Cat\'{o}lica de
Chile, Vicu\~{n}a Mackenna 4860, Santiago, Chile}
\address[emenendez]{Departamento de F\'{\i}sica, Facultad de Ciencias, Universidad
de Chile, Las Palmeras 3425 \~{N}u\~{n}oa, Santiago, Chile}

\date{\today}

\begin{abstract}
A modified version of the M{\o}ller-Plesset approach for obtaining the
correlation energy associated to a Hartree-Fock ground state is proposed.
The method is tested in a model of interacting fermions that allows for an
exact solution. Using up to third order terms improved results are obtained,
even in the limit of loosely bound particles.
\end{abstract}

\end{frontmatter}

The study of molecules and larger systems is seriously constrained by their
many body nature. Several approximation schemes have been devised over the
years, among which the Hartree-Fock (HF) method is one of the oldest and
most fruitful. Because it treats interactions in a mean field way particle
correlations are left out, however, a shortcoming that can limit severely the
validity of its results. One may improve over HF by treating correlations as
a perturbation. In the so-called M{\o }ller-Plesset method, a
Rayleigh-Schr\"{o}dinger perturbative expansion that is naturally suggested
by the same structure of the HF solution\cite{mp,grainer,march,szabo} is
adopted. This formalism reduces the correlation energy to an infinite series
in the perturbation, of which only the first few terms need  to be
computed in practice. This scheme has been used for a long time as a good starting
point to study correlation effects in molecular systems. However,
this method is useful only if the perturbation series
is rapidly convergent, which is not always true\cite{nobes87}.
Failures of the  M{\o }ller-Plesset method have been documented
even for small molecules \cite{nobes87,nobes91,ma92}.
In this paper we present a
variation of the M{\o }ller-Plesset approach that appears to give accurate
results in low order perturbation schemes even when particles are loosely
bound, such as in chemical bonds.

Consider the many-particles Hamiltonian
\begin{equation}
\mathcal{H}=\sum_{i}h(i)+\sum_{i,j>i}v(i,j),
\end{equation}
where $h(i)=h(\mathbf{r}_{i})$ is the sum of the one-particle kinetic energy
plus external potential energy, and $v(i,j)=v(\mathbf{r}_{i},\mathbf{r}_{j})$
is the two-particles interaction energy. The HF approximation to the ground
state of a system of identical fermions leads to a variational wave function
$\Phi $ in the form of a Slater determinant, constructed with one-particle
orbitals that satisfy the equations
\begin{equation}
\big[ h(i)+\sum_{b}(\mathcal{F}_{b}(i)-\mathcal{K}_{b}(i))\big] \phi
_{n}(i)=\epsilon _{n}\phi _{n}(i),
\end{equation}
where $\mathcal{F}_{b}$ and $\mathcal{K}_{b}$ are the Coulomb and exchange
operators, respectively. Here and in what follows summation over indices $%
a,b,c$ run over all occupied states, while \textsl{n} runs over all possible
states. The eigenvalues satisfy the relation
\begin{equation}
\epsilon _{n}=\langle n|h|n\rangle +\sum_{b}\langle nb\Vert nb\rangle ,
\label{spenergy}
\end{equation}
where
\begin{eqnarray}
\langle mn\Vert mn\rangle &=&\langle mn|mn\rangle -\langle mn|nm\rangle , \\
\langle mn|pq\rangle &=&\int \phi _{m}(1)^{*}\phi _{n}(2)^{*}v(1,2)\phi
_{p}(1)\phi _{q}(2)d1d2.
\end{eqnarray}
In the above expression 1 and 2 represent the one-electron variables of
coordinate and spin. The energy of the HF state is
\begin{eqnarray}
E_{HF} &=&\sum_{a}\langle a|h|a\rangle +\frac{1}{2}\sum_{a,b}\langle ab\Vert
ab\rangle \nonumber \\
&=&\sum_{a}\epsilon _{a}-\frac{1}{2}\sum_{a,b}\langle ab\Vert ab\rangle ,
\label{hfenergy}
\end{eqnarray}
where to obtain (\ref{hfenergy}) we have used Eq. (3). The operator
\begin{equation}
\mathcal{H}_{HF}=\sum_{n}\epsilon _{n}\hat{c}_{n}^{\dag }\hat{c}_{n}-\frac{1%
}{2}\sum_{a,b}\langle ab\Vert ab\rangle  \label{hfham}
\end{equation}
is then a natural choice for an effective HF hamiltonian, diagonal in the HF
orbitals and having as ground state the HF energy $E_{HF}.$ In
Eq. (\ref{hfham}) $\hat{c}_{n}^{\dag }$ ($\hat{c}_{n}$) is the creation
(annihilation) operator of a particle in the state $\phi _{n}.$ This
hamiltonian operator is the starting point to construct a perturbation
expansion for the correlation energy using as perturbation $V_{HF}=$ $%
\mathcal{H}-\mathcal{H}_{HF}$ \cite{szabo}.

A second choice for a hamiltonian is also possible, however. Equation (\ref
{hfenergy}) may be written in the form,
\begin{equation}
E_{HF}=\sum_{a}(\epsilon _{a}-\frac{1}{2}\sum_{b}\langle ab\Vert ab\rangle ),
\end{equation}
suggesting as an alternate hamiltonian
\begin{equation}
\mathcal{H}_{MHF}=\sum_{n}(\epsilon _{n}-\frac{1}{2}\sum_{b}\langle nb\Vert
nb\rangle )\hat{c}_{n}^{\dag }\hat{c}_{n},  \label{mhfham}
\end{equation}
also diagonal in the HF orbitals. The correlation energy can again be
described in terms of a perturbation, this time of the form $V_{MHF}=$ $%
\mathcal{H}-\mathcal{H}_{MHF}.$ Note that while Eq. (\ref{hfham}) involves
an overall constant, Eq. (\ref{mhfham}) substracts from each single particle
energy a different correction. Although both forms yield the same ground
state energy, the energy of excited states are different, affecting the
various orders contributions in a perturbation expansion. For instance, the
second order correction to the HF ground state energy has the form\cite{szabo}
\begin{equation}
E^{(2)}=\sum_{\Phi ^{\prime }}\frac{|\langle \Phi |\mathcal{H}|\Phi ^{\prime
}\rangle |^{2}}{E_{HF}-E_{\Phi ^{\prime }}}=\frac{1}{4}\sum_{abrs}\frac{%
|\langle \,ab\Vert \,rs\,\rangle |^{2}}{E_{HF}-E_{\Phi ^{\prime }}},
\label{soterm}
\end{equation}
where in the numerator we have used the fact that the excited state $\Phi
^{\prime }$ is orthogonal to the ground state. This is the first finite
correction in the Rayleigh-Schr\"{o}dinger perturbation expansion since for
either choice of hamiltonian the first order term vanishes$.$ Only doubly
excited states $\Phi ^{\prime }=\Phi _{ab}^{rs}$ ($\phi _{a}$ replaced by $%
\phi _{r}$ and $\phi _{b}$ replaced by $\phi _{s}$ in the Slater determinant
$\Phi ,$ with $\epsilon _{r},$ $\epsilon _{s}$ above the Fermi energy)
contribute\cite{szabo}. The choice of hamiltonian affects the excitation
energies in the denominator, which have the form
\begin{equation}
E_{HF}-E_{\Phi _{ab}^{rs}}=\left\{
\begin{array}{c}
\epsilon _{a}+\epsilon _{b}-\epsilon _{r}-\epsilon _{s},\quad \mathcal{H}%
_{0}=\mathcal{H}_{HF} \\
\tilde{\epsilon}_{a}+\tilde{\epsilon}_{b}-\tilde{\epsilon}_{r}-\tilde{%
\epsilon}_{s},\quad \mathcal{H}_{0}=\mathcal{H}_{MHF}
\end{array}
\right. ,  \label{denom}
\end{equation}
where $\tilde{\epsilon}_{n}=\epsilon _{n}-\frac{1}{2}\sum_{b}\langle nb\Vert
nb\rangle =\frac{1}{2}(\epsilon _{n}+\langle n|h|n\rangle )$. Replacing in
Eq. (\ref{soterm}) both forms clearly lead to different results. A similar
analysis of the third and higher order terms in the perturbation expansion
yields again different results. For example, in the third order correction,
besides the change of denominators in the standard M{\o }ller-Plesset
expression\cite{szabo}, an additional term $\Delta E^{(3)}$ appears, which
can be cast in the compact form, useful in computations,
\begin{eqnarray}
\lefteqn{\Delta E^{(3)}=-E^{(2)}}  \notag \\
&&-\frac{1}{4}\sum_{abrs}\frac{(h_{rr}+h_{ss}-h_{aa}-h_{bb})|\langle ab\Vert
rs\rangle |^{2}}{(\tilde{\epsilon}_{r}+\tilde{\epsilon}_{s}-\tilde{\epsilon}%
_{a}-\tilde{\epsilon}_{b})^{2}},  \label{eq:e3correc}
\end{eqnarray}
where $h_{nn}=\langle n|h|n\rangle $. The first element of Eq. (\ref
{eq:e3correc}) cancels the second order energy correction. However, the
second element of Eq. (\ref{eq:e3correc}) is also second order in the
interaction. Replacing Eqs.~(\ref{soterm}) and (\ref{denom}) in (\ref
{eq:e3correc}) and performing simple operations $\Delta E^{(3)}$ can be put
in a form that reveals explicitly the third order character of the
correction.

\begin{table}[tbh]
\caption{Ground state energies of a system of two harmonically confined spin
$1/2$ particles interacting through a harmonic potential of strength $k$.
Several approximations are included: Hartree-Fock (HF), M{\o }ller-Plesset
perturbation theory of orders $n=2,3$ (MP$n$), modified MP$n$ (MMP$n$), as
well as the exact values. Repulsive interactions are represented by negative values
of the elastic constant $k$.
\label{tab:ener}}
\begin{tabular}{ccccccc}
  \hline\hline\\
  $k$ & HF & MP2 & MP3 & MMP2 & MMP3 & Exact\\
  \hline
 -0.25 & 1.732 & 1.655 & 1.836 &  1.702 & 1.710 & 1.707 \\
 -0.24 & 1.744 & 1.681 & 1.803 &  1.717 & 1.724 & 1.721 \\
 -0.22 & 1.766 & 1.725 & 1.784 &  1.745 & 1.750 & 1.748 \\
 -0.20 & 1.789 & 1.760 & 1.791 &  1.772 & 1.776 & 1.775 \\
 -0.18 & 1.811 & 1.791 & 1.808 &  1.798 & 1.801 & 1.800 \\
 -0.16 & 1.833 & 1.819 & 1.828 &  1.823 & 1.825 & 1.825 \\
 -0.09 & 1.908 & 1.905 & 1.906 &  1.905 & 1.906 & 1.906 \\
 -0.04 & 1.960 & 1.959 & 1.959 &  1.959 & 1.959 & 1.959\\
 -0.01 & 1.990 & 1.990 & 1.990 &  1.990 & 1.990 & 1.990 \\
  0.00 & 2.000 & 2.000 & 2.000 &  2.000 & 2.000 & 2.000 \\
  0.04 & 2.040 & 2.039 & 2.039 &  2.039 & 2.039 & 2.039 \\
  0.16 & 2.154 & 2.150 & 2.149 &  2.149 & 2.149 & 2.149 \\
  0.36 & 2.332 & 2.319 & 2.314 &  2.316 & 2.313 & 2.311 \\
  0.64 & 2.561 & 2.534 & 2.522 &  2.525 & 2.516 & 2.510 \\
  1.00 & 2.829 & 2.784 & 2.762 &  2.767 & 2.749 & 2.732\\
\hline\hline\\
\end{tabular}
\end{table}

In order to assess the convenience of either formulation for obtaining
corrections due to correlations we have solved a system of two spin-1/2
particles in a harmonic potential, interacting through a harmonic force. The
reason for the choice is that this is an interacting system involving
identical fermions that may be solved exactly\cite{mosh}. The Hamiltonian is
\begin{equation}
H=\frac{1}{2}\left( -\Delta _{1}+r_{1}^{2}\right) +\frac{1}{2}\left( -\Delta
_{2}+r_{2}^{2}\right) +\frac{1}{2}k(\mathbf{r}_{1}-\mathbf{r}_{2})^{2},
\label{eq:2osc}
\end{equation}
where the coordinates and the energy are given in the oscillator units of
the confinement potential. The exact ground state energy is 
\begin{equation}
E_{0}=1+\sqrt{1+2k}.  \label{eq:exacten}
\end{equation}
The model contains the parameter $k$ that allows the study of attractive ($%
k>0 $) as well as repulsive ($k<0$) interactions. Equation (\ref{eq:exacten}%
) shows that $k=-0.5$ is the lowest value for which a bound state exists.

We have solved the problem in two dimensions, using a basis of
noninteracting harmonic oscillators eigenfunctions
\begin{equation}
\phi _{n_{x},n_{y}}(x,y)=\varphi _{n_{x}}(x)\varphi _{n_{y}}(y),\quad 0\le
n_{x}+n_{y}\le 5,  \label{set}
\end{equation}
where the $\varphi (x)$ are the usual one-dimensional harmonic oscillator
eigenfunctions. The set (\ref{set}) is an exact solution when $k=0$. The HF
solution is obtained by solving the self-consistent-field equation for
closed shell configurations\cite{szabo}.

\begin{figure}[tbh]
\includegraphics[width=10.5cm]{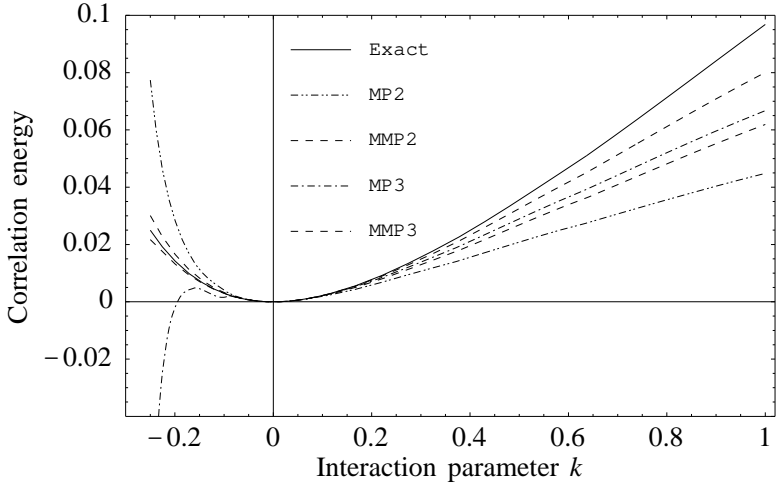}
\caption{Correlation energy for a system of two harmonically confined spin
1/2 particles interacting through a harmonic potential of strength $k$. The
exact result is included, as well as corrections up to second and third
order for the M{\o}ller-Plesset (MP) and modified M{\o}ller-Plesset (MMP)
choices of zeroth-order hamiltonian.}
\label{fig:osc}
\end{figure}

Table \ref{tab:ener} shows the ground state energy for the Hamiltonian
(\ref{eq:2osc}) calculated using the standard M{\o }ller-Plesset
perturbation theory of order $n=2,3$ (MP$n$), and our modified form, MMP$n$.
Figure \ref{fig:osc} shows the exact correlation energy $E_{corr}=E_{HF}-E$
for different values of $k$, together with results obtained for the two
choices of HF hamiltonian, in second and third order of perturbation theory.
Notice that MMP$n$ yields better results throughout. Notice also that the
usual MP$n$ fails badly in both orders of approximation when the system
becomes more loosely bound, as $k$ approaches the critical value -0.5. In
fact, convergence problems prevent solving the HF self-consistent equations
for $k$ beyond -0.25. By contrast, MMP$n$ continue to be good approximations
even in this range of $k$. This strongly suggests that the modified
perturbation theory may be more suitable in treating outer shells of bound
systems, such as electrons participating in chemical bonds. Investigation of
this ansatz is in progress.

A well known feature of HF theory is that the many body ground state energy
is not the bare sum of energies of filled single particle orbitals.
Interactions are counted twice and this overestimation is corrected for by
substracting the constant explicit in Eq. (\ref{spenergy}). The remarkable
improvement obtained over the usual M{\o }ller-Plesset approach in our test
example may be traced to the fact that this latter method does not correct
the single particle energies for such effect. In fact, the energy
denominator appearing in the perturbative corrections to all orders in such
case may grossly depart from the true two-particle excitation energy that
the numerator in the expression is weighting. By contrast, in the method
proposed here each self energy is corrected accordingly. It is hoped that
our results will stimulate the use of the proposed method in situations
where corrections to the Hartree-Fock approximations may be necessary.

Support from FONDECYT grants 1020829 and 7020829, and the Third World
Academy of Sciences, is gratefully acknowledged.

\end{document}